\documentclass[journal]{IEEEtran}

\usepackage{lipsum}
\usepackage{amsmath,mathtools,amsfonts,amssymb,amsthm}
\usepackage{xcolor}
\usepackage{svg}
\usepackage{cite}
\usepackage{epstopdf}
\usepackage{algorithm}
\usepackage[noend]{algpseudocode}  
 
\newcommand{\buses}{\mathcal{N}}
\newcommand{\lines}{\mathcal{L}}
\newcommand{\maximize}{\text{maximize}}
\newcommand{\Pc}{\overline{P^c}}
\newcommand{\Qc}{\overline{Q^c}}
\newcommand{\States}{\mathcal{S}}
\newcommand{\Actions}{\mathcal{A}}

\newcommand{\Rewards}{R}

\newcommand{\inserted}[1]{{\leavevmode\color{black}#1}}

\begin{document}

%%%%%%%%%%%%%%%%%%%%%%%%%%%%%%%%%%
% PAPER TITLE
%%%%%%%%%%%%%%%%%%%%%%%%%%%%%%%%%%
% Titles are generally capitalized except for words such as a, an, and, as, at, but, by, for, in, nor, of, on, or, the, to and up, which are usually not capitalized unless they are the first or last word of the title.
% Line breaks \\ can be used within to get better formatting as desired.
% e.g. \title{Bare Demo of IEEEtran.cls\\ for IEEE Journals}

\title{Fully Decentralized Reinforcement Learning-based Control of Photovoltaics in Distribution Grids for Joint Provision of Real and Reactive Power}

%%%%%%%%%%%%%%%%%%%%%%%%%%%%%%%%%%
% AUTHOR NAMES AND IEEE MEMBERSHIPS
\author{Rayan~El~Helou,~\IEEEmembership{Student Member,~IEEE,}
        Dileep~Kalathil,~\IEEEmembership{Senior~Member,~IEEE,}
        and~Le~Xie,~\IEEEmembership{Senior~Member,~IEEE}% <-this % stops a space
\thanks{The authors are with the Department of Electrical and Computer Engineering, Texas A\&M University, College Station, TX 77843, USA. e-mail: \{rayanelhelou, dileep.kalathil, le.xie\}@tamu.edu. This work was supported in part by NSF grants CRII-CPS-1850206, CAREER-EPCN-2045783, ECCS-1839616 and ECCS-2038963, and by Department of Energy Contract DE-EE0009031.}
}

%%%%%%%%%%%%%%%%%%%%%%%%%%%%%%%%%%
% PAPER HEADERS
% \markboth{TEMPORARY ONLY - Journal of \LaTeX\ Class Files,~Vol.~14, No.~8, August~2015}%
% {TEMPORARY ONLY - EL HELOU \MakeLowercase{\textit{et al.}}: Bare Demo of IEEEtran.cls for IEEE Journals}
%*******************************************************




\maketitle

%%%%%%%%%%%%%%%%%%%%%%%%%%%%%%%%%%
% ABSTRACT
\begin{abstract}
In this paper, we introduce a new framework to address the problem of voltage regulation in unbalanced distribution grids with deep photovoltaic penetration. \inserted{In this framework, both real and reactive power setpoints are explicitly controlled at each solar panel smart inverter, and the objective is to simultaneously minimize system-wide voltage deviation and maximize solar power output}. We formulate the problem as a Markov decision process with continuous action spaces and use proximal policy optimization, a reinforcement learning-based approach, to solve it, without the need for any forecast or explicit knowledge of network topology or line parameters. By representing the system in a quasi-steady state manner, and by carefully formulating the Markov decision process, we reduce the complexity of the problem and allow for fully decentralized (communication-free) policies, all of which make the trained policies much more practical and interpretable. Numerical simulations on a 240-node unbalanced distribution grid, based on a real network in Midwest U.S., are used to validate the proposed framework and reinforcement learning approach.
\end{abstract}

% Note that keywords are not normally used for peer review papers.

% \fi
\IEEEpeerreviewmaketitle

%%%%%%%%%%%%%%%%%%%%%%%%%%%%%%%%%%
% INTRODUCTION
%%%%%%%%%%%%%%%%%%%%%%%%%%%%%%%%%%
\section{Introduction}
% \IEEEPARstart{T}{he}

\IEEEPARstart{P}{hotovoltaic} (PV) smart inverter technology introduced in recent years enables solar panels to act as distributed energy resources (DERs) that can provide bi-directional reactive power support to electric power grid operations \cite{Farivar2012, Seuss_2015, Kekatos2015}. This support can be used to regulate local and system-wide voltages in distributed grids, and the IEEE Standard 1547-2018 \cite{IEEE_1547_2018} provides requirements on the use of such support. Voltage regulation is critical for network safety, both at the transmission and distribution levels.

\inserted{In the distribution grid, voltage regulation is usually controlled either through discrete switching (e.g. tap transformers, capacitor banks) or devices with continuous set points (e.g. PV inverters)}. Broadly speaking there are two categories of control and information structure to address the voltage regulation in distribution grids. The first category of solutions assumes complete or partial knowledge of system parameters and topology (such as a sample of references \cite{Su_2014, Zhang2015, Zhu2016, Lin2018, LiNa_Optimal, LiNa_limited_comm, ElHelou2020}). The second category of solutions primarily relies on observation data and do not explicitly use the knowledge of a physical model of the distribution network (such as a subset of references \cite{ADG, Xu2012, TwoTimescaleDeep, Brandon, SafeRL, ChangfuLi, WeiWang_2019}).

In the first set of solutions, control schemes are adopted based on assumed system models and on knowledge of line parameters. Given the broad set of references, we provide representative ones to illustrate some key directions of research. In \cite{Su_2014}, a multi-objective OPF problem is solved over a given unbalanced distribution grid model using Sequential Quadratic Programming (SQP), where the set of objectives include minimizing line losses, voltage deviation from nominal, voltage phase unbalance and power generation and curtailment costs. In \cite{Zhang2015}, a model-based voltage regulation problem is shown to be solvable by an equivalent Semi-Definite Programming (SDP) problem, and the sufficient conditions under which it can be solved as a result of convexifying the problem are outlined. In \cite{Zhu2016, Lin2018, LiNa_Optimal}, the distribution grid is modelled using the widely adopted linearized flow model, known as LinDistFlow, that assumes a tree (radial) grid structure and negligible line losses. In these papers, the same voltage regulation problem is solved, and an extension to these works are \cite{LiNa_limited_comm, ElHelou2020} in which limited or no communication between buses is needed and the same LinDistFlow model is adopted to provide theoretical guarantees on convergence and stability of the proposed control schemes.

\inserted{In the second set of solutions, reinforcement learning (RL) approaches are used to bypass the need to \textit{explicitly} model the system. {Nonetheless,} some form of a system model and a power flow solver is still needed to simulate the effects of actions on states, as it may not be feasible to learn  by directly interacting with a live power grid. However, knowledge of such a model or its parameters need not be \textit{explicitly} known by the solver of the RL problem, since optimal policies can be inferred solely by interacting with the simulation environment and receiving reward signals. This flexibility to system models is in part what makes RL approaches attractive, in contrast with the aforementioned approaches that rely heavily on a very specific class of system models. Furthermore, the set of objective functions commonly used in conventional control methods tend to be restrictive (e.g. quadratic), whereas in RL-based approaches, reward functions can be arbitrary and are usually designed to more directly reflect the user's underlying objective.}

In \cite{ADG}, for example, Batch RL is adopted to solve the optimal setting of voltage regulation transformers, where a virtual transitions generator is used to allow the RL agent to collect close-to-real samples, for learning, without jeopardizing real-time operation.
In \cite{Xu2012}, the Optimal Reactive Power Dispatch (ORPD) problem is solved using tabular Q-learning, where the objective is to minimize line losses, and actions include various forms of discrete reactive power re-dispatch, including switching control of tap transformers and of capacitor banks. Tabular Q-learning works well when there is a relatively small number of discrete states and actions, but suffers from the \textit{curse of dimensionality} and does not scale up well for larger systems.
In \cite{TwoTimescaleDeep}, \textit{deep} RL is used to optimize reactive power support over two timescales: one for discrete capacitor configuration and the other for continuous inverter setpoints. Here \textit{deep} refers to the representation of Q-value functions (used in Q-learning) in deep neural network form, as opposed to tabular form. This approach is well known as DQN \inserted{(Deep Q-learning)}. This is also applied in \cite{Brandon} to control voltages on the transmission level. While DQN uses neural networks to approximate value functions over continuous state spaces, it still relies on the assumption that the action space is discrete.
Policy gradients, which we review later in this paper, are an alternative set of RL approaches that enable continuous action spaces, and in \cite{SafeRL, ChangfuLi, WeiWang_2019}, policy gradients are used to solve Volt-VAR control problems where reactive power support is selected from a continuous set using a deep neural network that maps states directly to actions.

Such methods are inherently limited by physical constraints on reactive power support which are conventionally assumed to be uncontrollable. Here, we discuss the flexibility of such constraints and the value of relaxing them. \inserted{For example, conventional practices of maximum power-point tracking (MPPT) have been state-of-the-art, wherein each PV inverter is designed to extract the maximum real/active power from the solar panel. However, with a growing number of PV panels in the distribution grid, it becomes important to fully investigate the benefits and costs of always absorbing the maximum real power from the sun into the grid in real-time. By absorbing less real power, for instance, there is more room for reactive power support. In practice, real power curtailment at generators is performed either by solving model-based OPF problems, such as in \cite{Su_2014}, or by using a so-called VoltWatt (VW) droop controller, which requires carefully tuning the droop curves at each generator.}

In this paper, an RL approach is proposed for directly learning decentralized joint active and reactive power support. We illustrate a set of scenarios where instead of injecting all of the solar power into the network, it might rather be better to save or store the power and to inject it at a later time. Even in the absence of a storage system, under deep enough photovoltaic penetration, we
% \removed{find that it might surprisingly}
\inserted{discover that the RL agent surprisingly learns by itself that it might} be better to draw only parts of the available power, in order to avoid over-voltage, \inserted{especially} if there is an insufficient amount of reactive power resources available. This is accomplished by designing a reward function, used in the RL approach, that strikes a balance between voltage regulation and real power absorption.

\inserted{Despite modern advancements in deep RL, and in artificial intelligence and machine learning more broadly, the adoption rate of such tools in the power sector is still substantially low compared to other sectors. This is primarily due to the difficulty in interpreting deep neural networks, and due to the complexity of setting up processes that are required for the use of such tools in practice. These issues make it quite risky for a power grid operator to both trust those tools and to prefer them over conventional ones for operating over critical infrastructure. It is thus necessary for the adoption of RL by grid operators that the proposed approaches are heavily simplified and made much more interpretable, which is what we aim to contribute in this study.}

The key contributions of this paper are suggested as follows:
% \begin{enumerate}

%     \item \removed{A decentralized control policy architecture is proposed, that can be shown to train as well as, or better than, a centralized policy architecture in a continuous action space setting using reinforcement learning (RL).}
%     \item \removed{A parametrized reward function is proposed which enables the user to dictate the balance between minimization of voltage deviations from nominal and maximization of real power injection from solar panels, with results which illustrate that it might be better for a distribution grid to locally absorb a small fraction of real-time energy provided by solar panels, rather than inject it to the grid.}
    
% \end{enumerate}
\begin{enumerate}
    
    \item \inserted{Joint optimization of real and reactive power injection from the PV generation is formulated, with a parameterized reward function tailored for a multi-agent RL approach. Such parameterization is shown to facilitate more interpretable and easier to train deep RL policies, for a more user-friendly experience.}

    \item \inserted{A variant of PPO (Proximal Policy Optimization), a popular RL approach for handling continuous action spaces, is proposed for \textit{decentralized} settings, to dramatically simplify the search for optimal policies for relatively large systems. Not only can it be shown to train as well as or better than a centralized policy architecture, but it is much easier to implement than RL-based decentralized approaches in the existing body of literature.}
    
\end{enumerate}

% \removed{The RL agent observes voltages in the network and incrementally changes real and reactive power setpoints,}
\inserted{In our framework, an RL agent per bus observes voltages locally and incrementally updates real and reactive power setpoints at photovoltaic inverters,} similar to an integral droop controller (e.g. \cite{Katiraei2006}). However, it does not rely on any explicit knowledge of network topology or line parameters, and is fully decentralized, requiring minimal communication infrastructures for practical implementation. \inserted{This is illustrated in Fig. \ref{fig:agents_in_grid}, wherein agents are shown to be able to communicate with other agents or a central controller. The word \textit{fully} in the title of this paper refers to the fact that under our proposed framework, each agent in real-time (online) executes its policy locally without communicating with any other agent, not even any of its neighbors. During training though (while exploring and updating policies), observations are aggregated centrally so that all agents can adapt to one another. Since this is done on a much longer-term basis, the communication technologies used do not need to be as sophisticated for this purpose as they might be for an alternative online communication-based approach.

\begin{figure}[!ht]
\centering
\includegraphics[width=0.45\textwidth]{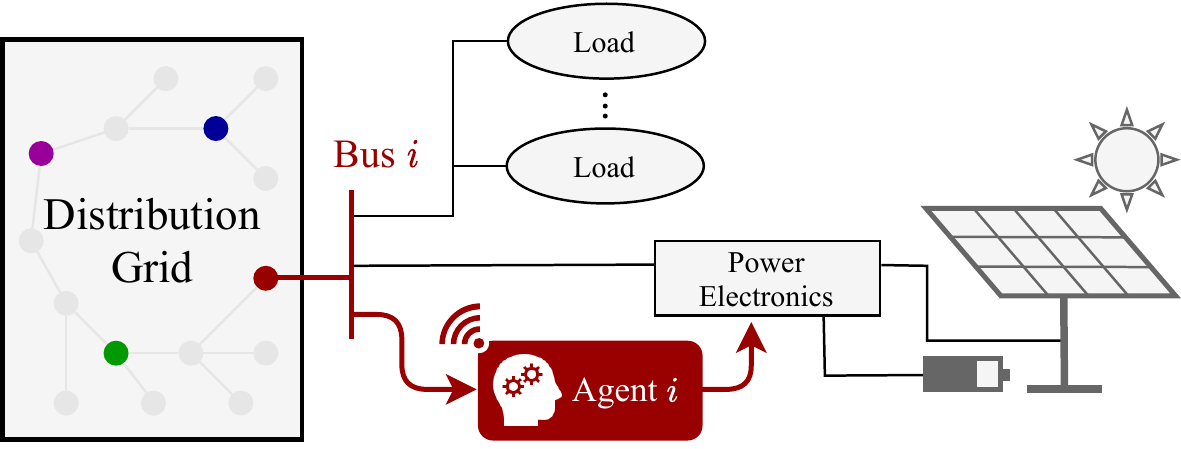}
\caption{\inserted{Every agent in the distribution grid controls a local photovoltaic system, with possible communication between agents. Note: each colored dot in the ``Distribution Grid" box refers to one agent.}}
\label{fig:agents_in_grid}
\end{figure}

Finally, the term ``load" in Fig. \ref{fig:agents_in_grid} is not restrictive. For example, it can include individual residential units or larger and more aggregate neighborhoods. In the former case, examples include houses with rooftop solar and the agent is controlling a few panels, and in the latter case, one bus could correspond to a solar farm and the agent would control many units simultaneously.}

The remainder of the paper is organized as follows. In Section \ref{Preliminaries}, the voltage regulation objective with joint real and reactive power compensation is formulated. In Section \ref{RL_section}, we provide a general review of Markov Decision Processes, and one specific to our problem in Section \ref{Voltage_Regulation_MDP}, with modifications to simplify the task. In Section \ref{policy_optimization}, centralized and decentralized policy architectures are proposed, and they are evaluated in Section \ref{numerical_simulation} with numerical simulations.

\section{Preliminaries}
\label{Preliminaries}

% \subsection{Three-phase Unbalanced Distribution Grid Model}
% \label{unbalanced_definition}

We consider a  three-phase balanced distribution network which consists of a set  $\buses =\{0,1,\hdots,N\}$ of buses and a set $\lines \subset\buses\times\buses$ of distribution lines connecting the buses. Bus $0$ represents the substation that acts as the single point  of connection to a bulk power grid. For each distribution line $(i, j ) \in  \lines$,   $y_{i,j}$ denotes its admittance.

% Let $\buses:=\{0,1,\hdots,N\}$ uniquely identify the set of buses in the network (one integer per bus), and zero is reserved for the substation bus. Similarly, let $\lines \subset\buses\times\buses$ uniquely identify the set of lines in the network, such that buses $i$ and $j$ are connected if and only if $(i,j)\in\lines$. For convention, $(i,i)\in\lines~\forall i\in\buses$.

The set of algebraic power flow equations that govern this three-phase balanced (single-phase equivalent) network are:
\begin{align}
    \label{eq:PF_balanced}
    P_i - \mathbf{i}Q_i = \widetilde{V}^*_i\sum_{j\in\buses}y_{ij}\widetilde{V}_j,
    \quad \forall i\in\buses,
\end{align}
where $P_i$ and $Q_i$ are the net injection of real and reactive power,  $\widetilde{V}_i$ is the complex phasor voltage at bus $i$. $\widetilde{V}^{*}_i$  denotes the complex conjugate of $\widetilde{V}_i$ and $\mathbf{i}:=\sqrt{-1}$.

To model a distribution grid that is not three-phase balanced, or \textit{unbalanced} for short, you may simply replace bus indices with phase indices in Eq. (\ref{eq:PF_balanced}), and $\buses$ with the set of all phases. With this, you can generalize over two-phase and single-phase buses, which are common in real distribution grids.

% \subsection{Voltage Regulation through both real and Reactive Power Compensation}
% \label{English Objective}

Let  $V_i$ denotes  the \textit{positive sequence} voltage magnitude at bus $i$.  At the substation bus, $V_0$ is fixed at 1.0 p.u. as it is modeled as an ideal voltage source.

% It is this voltage which we seek to regulate at each bus, with a desired setpoint of 1.0 p.u..

Let $\mathcal{C}\subset\buses$ be the set of buses that are equipped with solar panels and controllable smart inverters, and $n:=|\mathcal{C}|$. Let $P^c_i$ and $Q^c_i$ be the total real and reactive power, respectively, injected by the PV inverter at bus $i \in \mathcal{C}$. Each inverter has an apparent power capacity $S_i$, which limits $P^c_i$ and $Q^c_i$ as
\begin{align}
    \label{eq:inverter_constraints}
    (P^c_i)^2 + (Q^c_i)^2 \leq S_i^2, \hspace{0.4cm}  P^c_i \leq p^{\text{env}} \leq 0.9S_i,
\end{align}
where $p^{\text{env}}_i$ is the maximum amount of real power that can be drawn from the solar panel at a given moment in time. It depends on exogenous \textit{environmental} factors (irradiance, temperature, etc.), hence the superscript. The upper bound on this quantity is $0.9S_i$ since each inverter in the network is assumed to obey standard IEEE 1547-2018 \cite{IEEE_1547_2018}. We let this injected power be evenly distributed across all phases per bus.

\begin{figure}[!ht]
\centering
\includegraphics[width=0.45\textwidth]{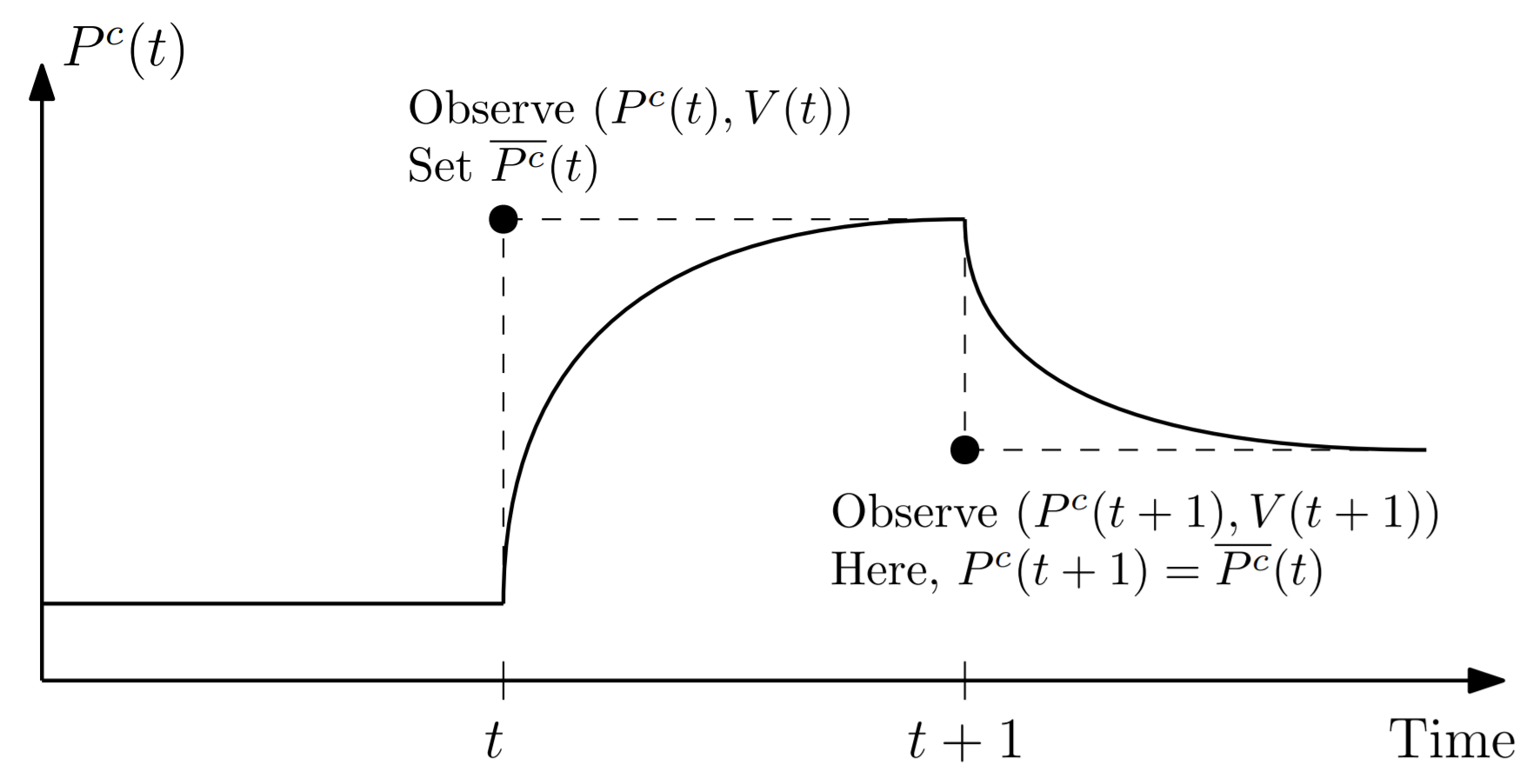}
\caption{Illustration of quasi-steady state behavior.}
\label{fig:quasi}
\end{figure}

Strictly speaking, if $P^c_i(t)$ is the actual real power injected by the inverter at time $t$, and $\overline{P^c_i}(t)$ is the setpoint, then those two cannot be equal at the same time. There is a small time delay ($\sim$ 10 ms, or less than one 60 Hz cycle) between when the setpoint is assigned and when the actual quantity tracks it. We let both the discrete time step and the tracking time be 10 ms.  This allows us to treat the system as a quasi-steady state system, as illustrated in Fig. \ref{fig:quasi}.

More precisely, net real and reactive power injections at bus $i$ at time $t$ can be expressed as $P_i(t) = P^c_i(t) - P^l_i(t)$ and $Q_i(t) = Q^c_i(t) - Q^l_i(t)$,  where $P^l_i(t)$ and  $Q^l_i(t)$ are the (uncontrollable) load consumption   at bus $i$ at time $t$. Due to the one time step delay between setpoint and actual injections, $P^c(t+1)=\overline{P^c}(t)$ and $Q^c(t+1)=\overline{Q^c}(t)$. Then, due to the quasi-steady state nature, $V(t+1) = f(\overline{P^c}(t), \overline{Q^c}(t))$, where $f(\cdot)$ represents the solution to the algebraic power flow equation given in \eqref{eq:PF_balanced}. 

We address the problem of voltage regulation in the distribution grid through joint real and reactive power control of PV inverter setpoints. The control objective is to track desired voltage levels while not wasting solar power in the process. The voltage regulation problem  can be formulated as follows: 
\begin{subequations}
    \begin{align}
        \begin{split}
            \label{eq:objective}
            \underset{\overline{P^c}, \overline{Q^c}}{\maximize} &\quad \mathbb{E}\left[\sum_{t=0}^{T}\sum_{i\in\mathcal{C}}R_{V_i(t)} + \mu_i R_{P^c_i(t)}\right],
        \end{split}\\
        \begin{split}
            \label{eq:net_P}
            \text{s.t. } &\quad P_i(t) = P^c_i(t) - P^l_i(t),
        \end{split}\\
        \begin{split}
            \label{eq:net_Q}
            &\quad Q_i(t) = Q^c_i(t) - Q^l_i(t),
        \end{split}\\
        \begin{split}
            &\quad \text{Eq. (\ref{eq:PF_balanced}), i.e. \textit{power flow}, } \forall t,
        \end{split}\\
        \begin{split}
            &\quad \text{Eq. (\ref{eq:inverter_constraints}), i.e. \textit{inverter constraints}, } \forall t,
        \end{split}\\
        \begin{split}
            \label{eq:R_V}
            & R_{V_i(t)} = \cfrac{1}{0.05}\min\left\{\delta - \left|1-V_i(t)\right|,0\right\},
        \end{split}\\
        \begin{split}
            \label{eq:R_P}
            & R_{P^c_i(t)} = \cfrac{P^c_i(t)}{0.9S_i},
        \end{split}
    \end{align}
\end{subequations}
where $\mu_{i}$'s and $\delta$ are positive scalars. Unlike with $P^c$ and $Q^c$ (controllable inverter), $P^l$ and $Q^l$ (uncontrollable load) are generally not evenly distributed across all individual phases. For clarity, subscript $i$ refers to $i^\text{th}$ bus index, and terms where the subscript is dropped refer to all buses collectively.

\begin{figure}[!ht]
\centering
\includegraphics[width=0.49\textwidth]{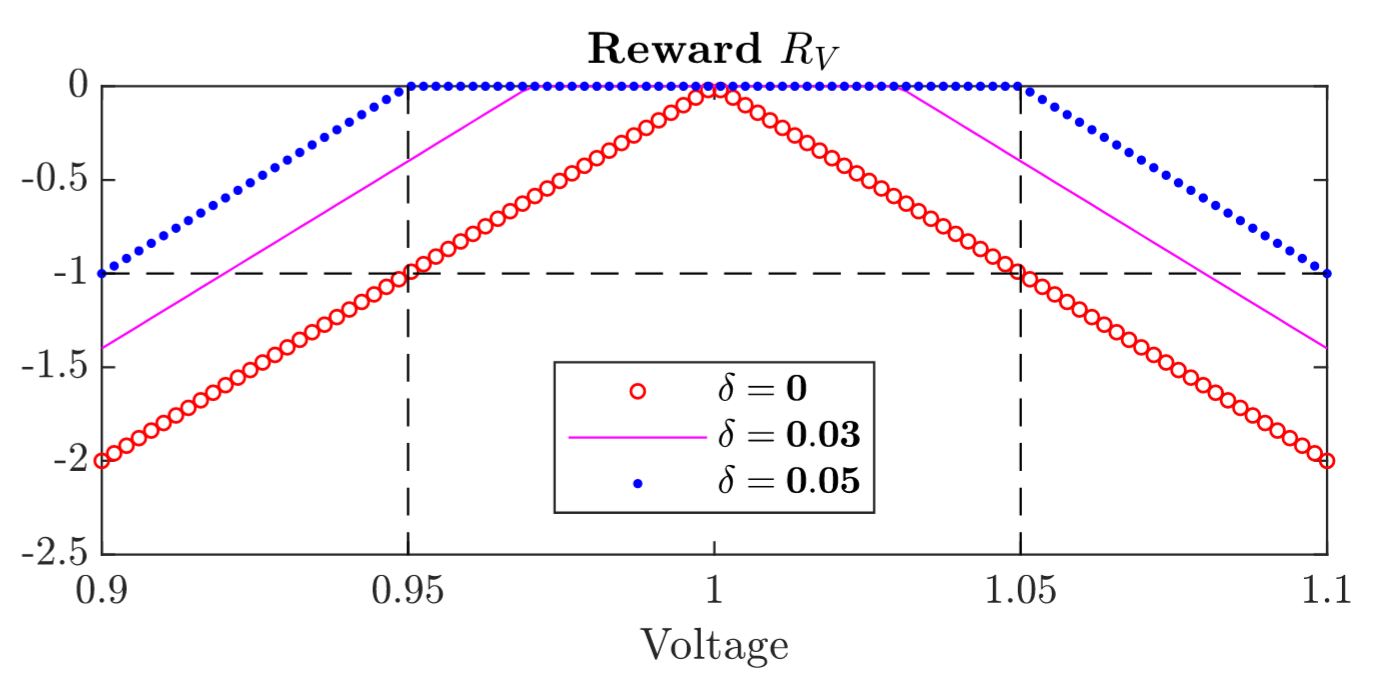}
\caption{Voltage deviation reward function $R_V$ for different $\delta$ (see Eq. (\ref{eq:R_V})).}
\label{fig:R_V}
\end{figure}

Voltage deviation (from nominal 1.0 p.u.) at each bus is considered acceptable if it is kept within some user-defined $\delta$. Deviations greater than this are assigned negative rewards, as depicted in Fig. \ref{fig:R_V}, to signify an undesirable voltage profile. The reward term $R_V$ in \eqref{eq:R_V} quantifies this criteria.

In conventional voltage regulation schemes, reactive power injection or consumption is used to alleviate voltage deviation problems. Reactive power support is limited by physical constraints. For example in the case of PV inverters, as expressed in Eq. (\ref{eq:inverter_constraints}), real power injection by the solar panel directly limits available reactive power. In our proposed objective, we not only consider reactive power but real power provision as well as a decision variable to add a degree of freedom and to relax constraints on reactive power support if needed. Since we seek to extract as much real power as possible from the solar panel, physically bounded by $0.9S_i$ as expressed in Eq. (\ref{eq:inverter_constraints}), we assign a positive reward  to more power drawn from the solar panels, as quantified in \eqref{eq:R_P} and \eqref{eq:objective}.

Under this framework with joint provision of real and reactive power, the user (e.g. utility) selects parameter $\mu$. This parameter acts as a balancing term, between voltage deviation minimization and solar production maximization, considering the fact that over-injection of power leads directly both to over-voltage and to tighter constraints on reactive power. A value for $\mu$ chosen too high yields a control scheme equivalent to conventional maximum power point tracking (MPPT) control with limited reactive power support. On the other hand, as $\mu$ approaches zero, much less real power is likely to be drawn from solar panels by the optimal controller. In section \ref{numerical_simulation}, we demonstrate a balanced choice of $\mu$.

% Note that not every bus in $\mathcal{C}$ is equipped with a storage system, but if one is available at some bus $i\in\mathcal{C}$, it is reasonable to simply set $\mu_i$ to zero to allow all real power to be redirected away from the grid, back to the same bus, and into the storage system. Conversely, when there is no storage system at the bus, a non-zero $\mu_i$ is needed to capitalize on available `free' resources.

The following assumptions are made about variables that are not explicitly controlled:
\begin{itemize}
    \item While provided to the simulator, neither network topology nor line parameters are used by the controller at any time during training or execution. That is, no a priori knowledge of such values is needed.
    \item No load or solar forecasting is made available to the controller, neither upon training nor during execution.
    % \item The maximum amount of real power the controller can draw from the solar panel, at any point in time, is not known unless the control setpoint is set to that amount or greater.
    \item Net load at a control bus is measured by the controller before supplying a setpoint to the solar panel inverter.
\end{itemize}
The voltage regulation problem formulated in this section is posed as a Markov Decision Process (MDP) in Section \ref{Voltage_Regulation_MDP}, but first, in Section \ref{RL_section}, MDP terminology and formalism is introduced for a more general class of control problems.

%%%%%%%%%%%%%%%%%%%%%%%%%%%%%%%%%%
% REINFORCEMENT LEARNING FRAMEWORK
%%%%%%%%%%%%%%%%%%%%%%%%%%%%%%%%%%
\section{Markov Decision Process and Reinforcement Learning}
\label{RL_section}
In this section, we give a brief review of the Markov Decision Processes (MDP) terminology. We will also review a specific RL algorithm, called Proximal Policy Optimization (PPO) \cite{PPO}, which we use to solve the voltage regulation problem. 

MDP is a standard  formalism for modeling and solving  control problems. The goal is to solve sequential decision making (control) problems  where the control actions can influence the evolution of the state of the system. An MDP can be defined as a four-tuple $(\States,\Actions,\Bar{P},\Rewards)$, where $\States$ is the state space and $\Actions$ is the action space. $\Bar{P}(s'|s,a)$ is the probability of transitioning from state $s$ to $s'$ upon taking action $a$, and $R(s)$ is the reward collected at this transition. We consider a finite horizon MDP setting with horizon (episode) length of $T$. We assume that the rewards depend only on the state, not on the actions. A control policy $\pi: \mathcal{S} \rightarrow \mathcal{A}$ specifies the control action to take in each possible state. The performance of a policy is measured using the metric of value of a policy, $\Bar{V}_{\pi}$, defined as,
\begin{align}
    \label{eq:value_fcn}
    \Bar{V}_{\pi}(s) = \mathbb{E}_{\pi} \left[\sum^{T-1}_{t=0} R(s(t)) | s(0) = s \right],
\end{align}
where, $s(t+1) \sim \Bar{P}(\cdot|s(t),a(t)), a(t) = \pi(s(t)),$ and $s(t)$ is the state of the system at time $t$ and $a(t)$  is the action taken at time $t$. The goal is to find the optimal policy $\pi^{*}$ that achieves the maximum value, i.e, $\pi^{*} = \arg \max_{\pi} \Bar{V}_{\pi}$. The corresponding value function, $\bar{V}^{*} = \bar{V}_{\pi^{*}}$, is called the optimal value function. $\pi^{*}$ and $\bar{V}^{*}$ satisfy the Bellman equation, 
\begin{align}
    \pi^*(s) = \underset{a\in\Actions}{\text{argmax}}\left[R(s) + \sum_{s'\in\States} \bar{P}(s'|s,a) \bar{V}^{*}(s')\right]
\end{align}

When the system model $\bar{P}$ is known, the optimal policy and value function can be computed using dynamic programming \cite{sutton1998}. However, in most real-world applications, the  system model is either unknown or extremely difficult to model. In the voltage regulation problem, the line parameters and/or the topology of the network may not be known a priori. Even if they were known, it would still be difficult to model the effect of actions on states in a feed-forward fashion, due to the algebraic nature of their relationship.  In such scenarios, the optimal policy has to be learned from sequential state/reward observations by interacting with an environment, which in this paper is a simulation environment as described in Section \ref{Voltage_Regulation_MDP}. Reinforcement learning is the approach for computing the optimal policy for an MDP when the model is unknown.

Policy gradient algorithms are  a popular class of RL algorithms. In a policy gradient algorithm, we represent the policy as $\pi_{\theta}$, where $\theta$ denotes parameters of the neural network used to represent the policy. Let $J(\theta) = \mathbb{E}_{s}[\bar{V}_{\pi_{\theta}}(s)],$ where the expectation is w.r.t. to a given initial state distribution.  The goal is to find the optimal parameter $\theta^{*} = \arg \max_{\theta} J(\theta)$. This is achieved by implementing a gradient descent update, $\theta_{k+1} = \theta_{k} + \alpha_{k} \nabla J(\theta_{k})$, where $\alpha_{k}$ is the learning rate. The gradient $\nabla J(\theta)$ is given by the celebrated policy gradient theorem \cite{sutton1998} as $\nabla J(\theta) = \mathbb{E}_{\pi_{\theta}}[\Bar{Q}_{\pi_{\theta}}(s,a) \nabla \log \pi_{\theta}(s,a)]$, where expectation is w.r.t. the state and action distribution realized by following the policy $\pi_{\theta}$. Here $\Bar{Q}_{\pi_{\theta}}$ is the Q-value function corresponding to the policy $\pi_{\theta}$. Often, the Q-value function is represented using a neural network of its own (different from the one used for policy representation). The neural network which represents the policy is called the actor network and that which represents the value function is called the critic network. The terminology is due to the fact that the policy network determines actions given observations and the value network provides the `critic' feedback to update the policy parameter, as clear from the expression for $\nabla J(\theta)$. This class of algorithms is also called actor-critic algorithms. The goal is to incrementally update the parameters of both networks in such a way that they converge to yield parameters corresponding to optimal policy and Q-value functions.
 
Trust region policy optimization (TRPO) \cite{TRPO} is a recent variant of policy gradient algorithms. For improving the sample efficiency and ensuring reliable convergence, TRPO modifies the policy update as
\begin{align}
    \theta_{k+1} &= \underset{\theta}{\text{argmax}}~ \mathbb{E}_{\theta_{k}} \left[\frac{\pi_{\theta}(s,a)}{\pi_{\theta_{k}}(s,a)} \Bar{Q}_{\pi_{\theta_{k}}}(s, a)\right] \\
    &\text{s.t.}~~ \mathbb{E}\left[D_{\text{KL}}(\pi_{\theta_{k}}(s, \cdot), \pi_{\theta}(s, \cdot)) \right] \leq d,
\end{align}
where, $D_{\text{KL}}(\cdot, \cdot)$ is the Kullback-Leibler divergence between two policies, and $\pi_{\theta}(s,a)$ denotes probability of selecting action $a$ given state $s$. Constant $d$ is a user-defined threshold.

Proximal policy optimization (PPO) algorithm \cite{PPO} builds upon the TRPO framework by modifying the objective function and optimization update which enables improved data efficiency and easier implementation since the KL divergence constraint is dropped and the objective is clipped (or saturated) as described in \cite{PPO}. Clipping of the objective is performed to discourage the optimizer from over-updating $\theta$. We adapt this state-of-the-art algorithm to a decentralized setting to solve the voltage regulation problem we consider.

%%%%%%%%%%%%%%%%%%%%%%%%%%%%%%%%%%
% VOLTAGE REGULATION FORMALIZED AS AN MDP
%%%%%%%%%%%%%%%%%%%%%%%%%%%%%%%%%%
\section{Voltage Regulation  as an RL Problem}
\label{Voltage_Regulation_MDP}
We use the MDP formalism to model the voltage regulation problem presented in Section \ref{Preliminaries}. As a reminder, $n:=|\mathcal{C}|$.

\subsubsection{State space}
The state space $\States\subset\mathbb{R}^{2n}$ is the set of real power injections and voltage measurement at all controllable buses. For convenience, each state $s\in\States$ is defined as an affine transformation of those measurements. More precisely, the state of the system at time $t$, $s(t)$, is given by
\begin{align}
    \label{eq:s_PV}
    s(t) := (&s^P_1(t),~s^V_1(t),~\cdots,~s^P_n(t),~s^V_n(t)),  \\
    \text{where,}~ &s^P_i(t) \gets \cfrac{P^c_i(t)}{0.9S_i}-1,~~ s^V_i(t) \gets \cfrac{1-V_i(t)}{0.05}. \nonumber 
\end{align}
So, when $P^c_i(t)$ is its maximum allowable value, $s^P_i(t)$ is zero. Also, when the voltage is equal to the nominal value,  $s^V_i(t)$ is zero. Thus, ideal scenarios correspond to the state value zero, and critical scenarios correspond to magnitudes of order one or less, assuming critical voltages exceed $1\pm0.05$. This scaling helps to initialize and train the RL algorithm.

\subsubsection{Action space}
Given voltage measurements at every bus in $\mathcal{C}$, we can choose two approaches to determine $\Pc$ and $\Qc$: 1) Directly determine optimal real and reactive power setpoints, i.e. $\left(\Pc,\Qc\right)$, by algebraically tying to voltage, or 2) change setpoints incrementally, i.e. $\left(\Delta\Pc,\Delta\Qc\right)$, similar to an integral controller. The first approach requires the design and memorization of a highly non-linear function that is likely dependant on system operating conditions. Due to the lack of tracking in this approach, forecasting would be required to respond to different operating conditions. The second approach, on the other hand, enables tracking a desired state in a simple and incremental way \cite{ElHelou2020}. We use the second approach in this paper.

The action space $\Actions \in [-1,1]^{2n}$ is the set of possible \textit{scaled increments} in real and reactive power setpoints. The action at time $t$, $a(t)$, is given by $a(t) = ( a^P_1,~a^Q_1, \cdots,~a^P_n,~a^Q_n)$  and the increment to those setpoints are defined respectively as $\Delta\overline{P^c_i} = a^P_i\cdot\Delta_{\max}^P$ and $\Delta\overline{Q^c_i} = a^Q_i\cdot\Delta_{\max}^Q$. Here $\Delta_{\max}$ explicitly limit the size of actual (as opposed to scaled) increments $\left(\Delta\Pc,\Delta\Qc\right)$.

\subsubsection{Transition Model}
We assume that \textit{next states} are obtained by interaction either with a real-world distribution grid or with a simulator, such as OpenDSS \cite{OpenDSS}. In the case of a simulator, provided changes in loads $(P^l,~Q^l)$ and current state and action based on $\States$ and $\Actions$, next states can be computed directly. For example, actions are mapped to states using OpenDSS as follows:
\begin{align}
    V(t+1) = \text{OpenDSS}(\Pc(t),\Qc(t))
\end{align}
We choose to use OpenDSS simulator for two main reasons: (a) It can solve power flow for unbalanced distribution grids, and (b) one can directly interact with it using Python, where RL methods are easier to implement.

\subsubsection{Reward function}
System-wide reward at every time step is obtained as follows:
\begin{align}
    R_t = \cfrac{1}{n}\sum_{i\in\mathcal{C}}R_{V_i(t)} + \mu_i R_{P^c_i(t)}
\end{align}
where $R_{V_i(t)}$ and $R_{P^c_i(t)}$ are defined in Eq. (\ref{eq:R_P},\ref{eq:R_V}).

Note that the state and action spaces have been defined in such a way that each element ranges from $-1$ to $1$, with an exception where the voltage-related state may exceed $\pm 1$ if the p.u. voltage exceeds $1\pm 0.05$ under abnormal conditions. This is a suitable choice for training an RL policy as it allows for initialization and adjustment of policy parameters $\theta$ in a standard way by exploiting the  state of the art algorithms (most of  which requires that state and action spaces be a \textit{box} inside $\pm 1$ along all dimensions).

% \subsection{RL agent nested in integral controller}
\label{nested_RL_integral}
Based on the definition of action space, the RL agent seeks to learn the \textit{magnitude and direction} in which to incrementally change the setpoints, for every starting state. This raises the question: what information does the agent need to guide this action? The state defined in Eq. (\ref{eq:s_PV}) has the following advantage: If both the voltage term and the power term are zero (i.e. maximum power drawn and nominal voltage), then the scenario is ideal and no \textit{extra} injection is needed. If the load changes in the system, though, a simple amendment to the RL controller is needed:
\begin{equation}
    \label{eq:nested}
    \begin{aligned}
        \Delta\Pc &\gets \Delta_{max}^P\cdot a_P +\Delta P^l \\
        \Delta\Qc &\gets \Delta_{max}^Q\cdot a_Q +\Delta Q^l \\
    \end{aligned}
\end{equation}
where $a_P$ and $a_Q$ (both in $[-1,+1]$) are determined by the RL agent's zero-centered policy $\pi$, and $\left(\Delta P^l, \Delta Q^l\right)$ is the observed change in load at the controllable buses.

The strategy adopted in Eq. (\ref{eq:nested}) is termed an integral controller since setpoint $(\Pc,\Qc)$ behaves as a discrete-time integrator of changes in operating conditions. Moreover, this controller tracks the state to zero in steady state, within resource limits, since all terms in Eq. (\ref{eq:nested}) go to zero if $s=0$. Under scarcity of resources, one or more of the state terms in Eq. (\ref{eq:s_PV}) will be non-zero, which calls for a balance between \textit{maximum power point tracking} and \textit{voltage regulation}.

Note that state-tracking incremental setpoint changes are bounded by $\Delta_{max}^P$ and $\Delta_{max}^Q$ to limit fluctuations. These values are chosen heuristically as $0.09S_i$ and $0.2S_i$ respectively since those are one-tenth of the maximum possible jumps in setpoints $\Pc$ and $\Qc$.

%%%%%%%%%%%%%%%%%%%%%%%%%%%%%%%%%%
% POLICY ARCHITECTURE
%%%%%%%%%%%%%%%%%%%%%%%%%%%%%%%%%%
\section{Control Policy Architecture and Optimization}
\label{policy_optimization}
\begin{figure}[!ht]
\centering
\includegraphics[width=0.45\textwidth]{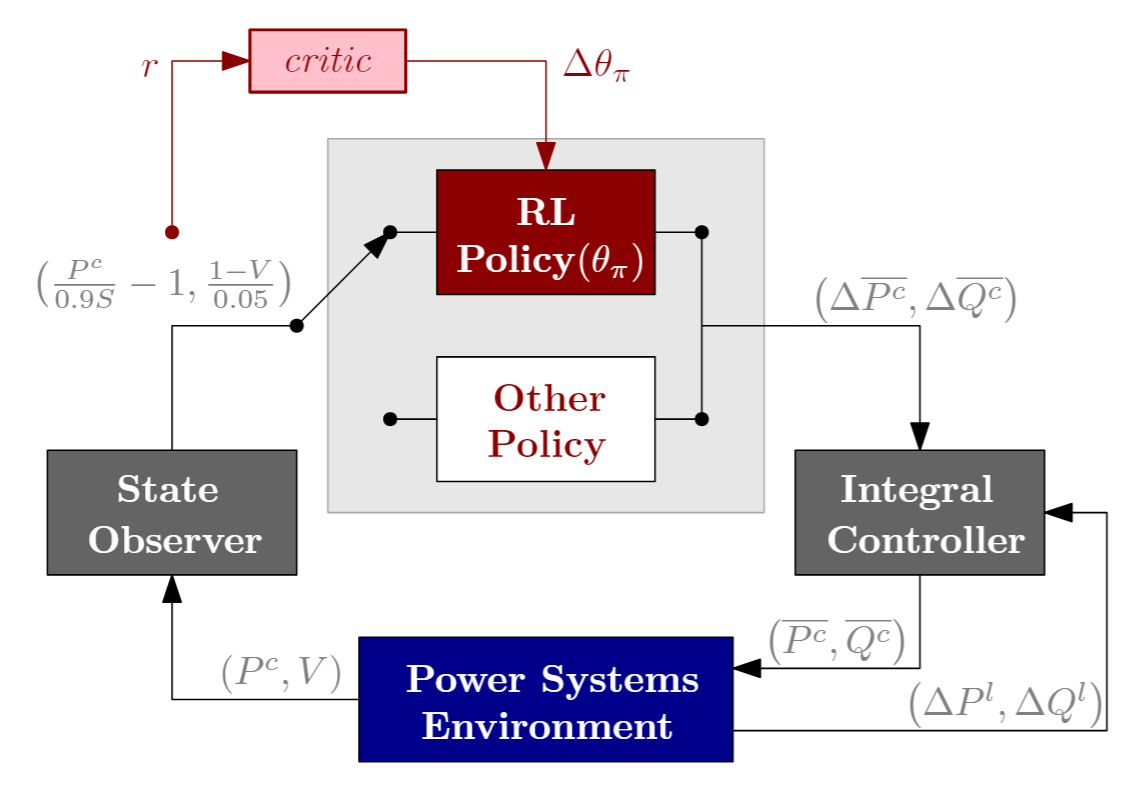}
\caption{RL-based Control Architecture.}
\label{fig:framework}
\end{figure}

In this section, we present the design and architecture of our RL algorithm for voltage regulation. We build on the PPO \cite{PPO}  algorithm and extend it to a decentralized setting. Fig. \ref{fig:framework} summarizes the RL-based control policy architecture we propose. \textit{State Observer} refers to Eq. (\ref{eq:s_PV}) and \textit{Integral Controller} refers to Eq. (\ref{eq:nested}). $\Delta\theta_\pi$ refers to changes in policy $\pi$ determined by the PPO algorithm. The policy and critic network architectures used to implement PPO are detailed in the remainder of this section.

The standard PPO algorithm assumes that there is a single centralized agent which fully observes the state of the system and can take any control action. However, this centralized control policy may not be feasible for the voltage regulation problem we consider. Firstly, the RL algorithm may not be able to scale to a large network with many nodes. Secondly, even if a centralized training of an RL algorithm is feasible,  the implementation of such a centralized control policy in the real-world system may not be possible due to the communication infrastructure needed. We propose a decentralized RL algorithm that overcomes these challenges.

Recall that $s_i$ and $a_i$ refer local state and action (at bus $i$) respectively. As defined in Eq. (\ref{eq:s_PV}), each state $s_i$ contains two terms per bus, relating to real power and voltage measurements. Similarly, action $a_i$ was defined in such a way that it also contains two terms per bus, relating to changes in real and reactive power setpoints. Our goal is to find the optimal policy parameter $\theta^*_i$ for each bus $i$ that maps local state $s_{i}(t)$ to local control action $a_{i}(t)$, i.e. $a_{i}(t) = \pi_{\theta^{*}_{i}}(s_i(t))$, in an optimal way. The objective is to maximize the cumulative global (system-wide) reward.

To replace the centralized policy, used in standard PPO, with a decentralized policy, we propose a neural network architecture for $\pi$ that connects input to output only at the same bus, rendering it equivalent to a decentralized controller, to compete with conventional methods. That is, there are $n$ neural networks in parallel, each with just 2 inputs and 2 outputs. Each of those smaller networks is parametrized by a group of weights and biases, denoted collectively as $\theta_i$, and notation $\pi_{\theta_i}$ is shortened to $\pi_i$. This time, the PPO algorithm optimizes over $\theta_1,\hdots,\theta_n$, in search of optimal policies $\pi_1,\hdots,\pi_n$, where
\begin{align}
    a_i \gets \pi_i(s_i) \quad \forall i \in \mathcal{C}
\end{align}

Note that the only difference between this case and the centralized case (optimizing over $\theta$), is that here we enforce the strict rule that all neural network weights connecting states at bus $i$ to actions at bus $j$ are fixed at zero iff $i\neq j$. One can also modify this architecture by replacing the condition $i\neq j$ with $(i,j)\notin\mathcal{L}$, if the desired setup involves neighboring buses communicating with one another. For training and implementation purposes, we perform orthogonal initialization on neural network weights for all policies and assign very small initial values to those in the last layer to prevent instability in the feedback controller.

In PPO, value function $\Bar{V}$ (see Eq. (\ref{eq:value_fcn})) is also trained at every iteration when the policy is trained. Even though we desire a decentralized control setting, actor policies are not trained to optimize over local reward maximization, rather they are trained to maximize global (system-wide) reward. For this reason, instead of $n$ value functions ($\Bar{V}_i$ for each $i\in\mathcal{C}$), there is a single value function (critic) for all actors combined, and this function's argument is the system-wide state. In contrast, Multi-Agent Deep Deterministic Policy Gradients (MADDPG) \cite{MADDPG}, a method that also employs multiple policy networks in a policy gradient approach, expects each `agent' not only to observe and act locally (with its own policy network), but to also have its own critic network (value function estimate) and to estimate the policy network of others. In that paper, each agent is assumed to have their own reward function or objective. However, since there's always a global objective (single reward signal), then only one critic function suffices, and the policy networks train simultaneously to adapt to each other.

Due to the way the policy architecture was chosen, there is still a single policy function $\pi$ from a standard PPO algorithm perspective, with $2n$ inputs and $2n$ outputs in this formulation. Since there is a single value function, the PPO agent is trained as usual, without any explicit changes to the algorithm, other than the aforementioned restriction that the policy network connects input to output only at the same bus. To implement this, the optimizer (e.g. Adam optimizer in PyTorch) is told to ignore the weights (initialized and left at zero) in the policy network which link actions at one bus to states at another.

\begin{algorithm*}
\caption{\inserted{One Episode of Interacting With the Distribution Grid}}\label{alg:algo_episode}
\inserted{
\begin{algorithmic}[1]
\renewcommand{\algorithmicloop}{\textbf{function }}
\Loop \texttt{RunEpisode}(\texttt{in\_training}, $\mathcal{D}$) \Comment{\texttt{in\_training} is True or False, $\mathcal{D}$ is experience buffer}
\State Specify load \& solar scenario (random \& time of day).
\If{\texttt{in\_training}}
    \State Set agent policies to sample actions stochastically
\Else 
    \State Set agent policies to sample actions deterministically
\EndIf
\While{episode not done}
    \For{each controllable bus $i$ in grid}
        \State Observe local voltage and active power measurements.
        \State Use policy $\pi_i$ to sample changes to active \& reactive power setpoints at local PV panels.
    \EndFor
    \State Collect rewards and append actions, observations, \& rewards to $\mathcal{D}$. \Comment{Do this only if \texttt{in\_training}}
\EndWhile
\EndLoop
\end{algorithmic}
}
\end{algorithm*}

\begin{algorithm*}
\caption{\inserted{Training Decentralized PPO for Distribution Grids}}\label{alg:algo_train}
\inserted{
\begin{algorithmic}[1]
\State Initialize the following:
\begin{itemize}
    \item policy neural network, $\pi_i$ for each bus $i$
    \item value neural network, $\bar{V}$
    \item PPO hyper-parameters
    \item experience buffer, $\mathcal{D}$ \Comment{memory of actions, observations and rewards used in PPO}
\end{itemize}
\State \texttt{in\_training} $\gets$ True
\While{number of episodes less than threshold}
    \For{each episode}
        \State \texttt{RunEpisode}(\texttt{in\_training}, $\mathcal{D}$) \Comment{Defined in Algorithm \ref{alg:algo_episode}}
    \EndFor
    \renewcommand{\algorithmicwhile}{\textbf{every}}
    \While{pre-specified number of episodes}
        \State Use $\mathcal{D}$ to update policy and value networks according to PPO.
    \EndWhile
\EndWhile
\end{algorithmic}
}
\end{algorithm*}

%%%%%%%%%%%%%%%%%%%%%%%%%%%%%%%%%%
% NUMERICAL SIMULATION
%%%%%%%%%%%%%%%%%%%%%%%%%%%%%%%%%%
\section{Numerical Simulation}
\label{numerical_simulation}
In this section, we apply the proposed policy architecture and use PPO to solve the MDP. Numerical simulations are conducted on a 240-node distribution grid (see Fig. \ref{fig:grid}) using OpenDSS to solve unbalanced power flow, as described in Section \ref{Voltage_Regulation_MDP}. All parameters associated with this network are obtained from real line parameters and real load data, based on an anonymous distribution grid in Midwest U.S. \cite{IowaState}. Experiment details (e.g. software and hardware details) are found in the Appendix \ref{Appendix_Experiment_Details}.

Based on numerical simulations, as shown in Fig. \ref{fig:rewards}, we have found that the decentralized agent is more sample efficient and trains with fewer fluctuations and variance in episodic rewards over the learning process. On the other hand, the centralized agent takes a bit less computation time (about 20\% less) per iteration, yet takes more iterations to converge.

\subsection{Simulation Setup}
\begin{figure}[!ht]
\centering
\includegraphics[width=0.45\textwidth]{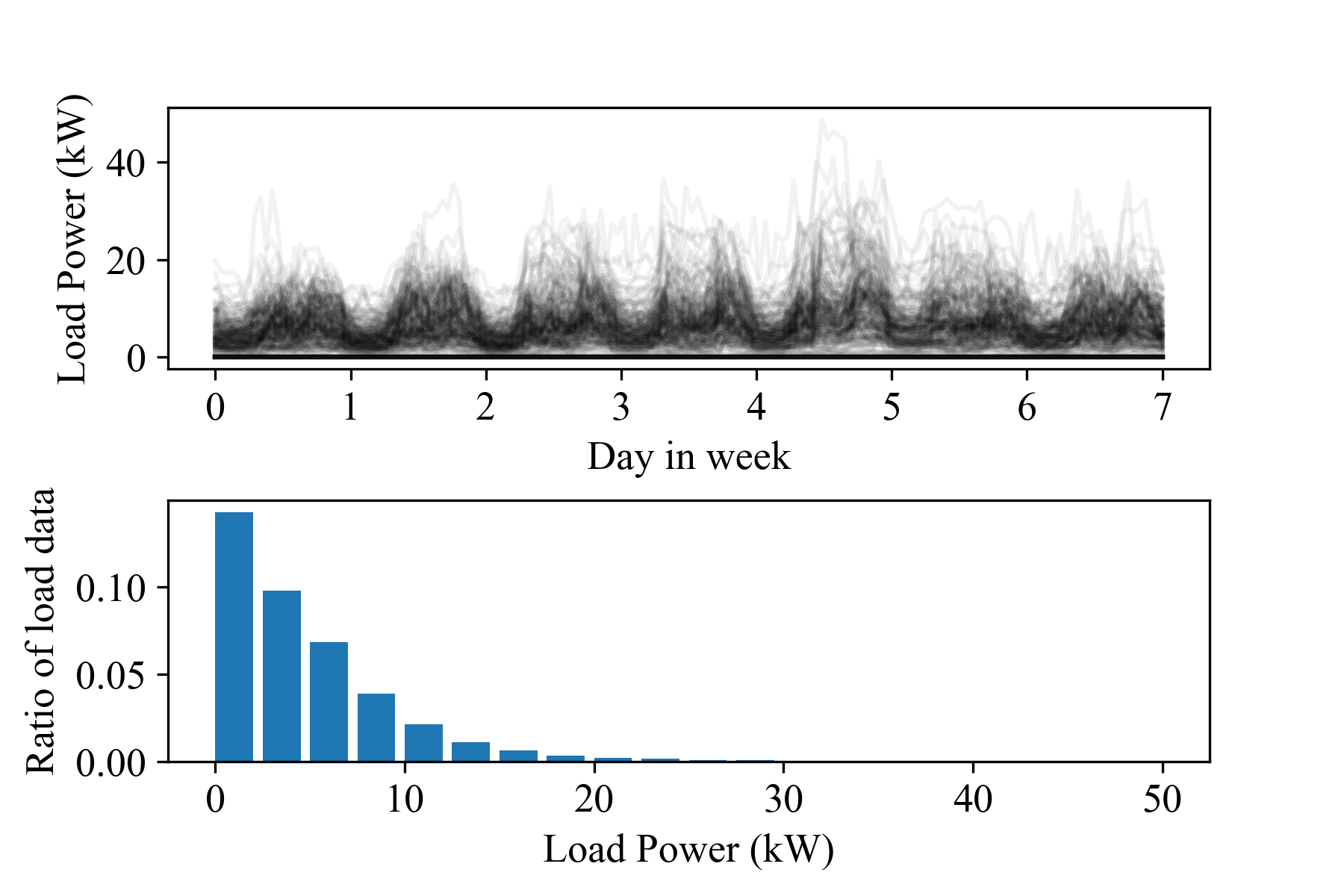}
\caption{\inserted{Example of real load data from grid located in the Midwest U.S. \cite{IowaState}. (Top) about 200 load profiles in the first week of January 2017. (Bottom) load distribution over the entire year of 2017.}}
\label{fig:load_stat}
\end{figure}

\begin{figure}[!ht]
\centering
\includegraphics[width=0.45\textwidth]{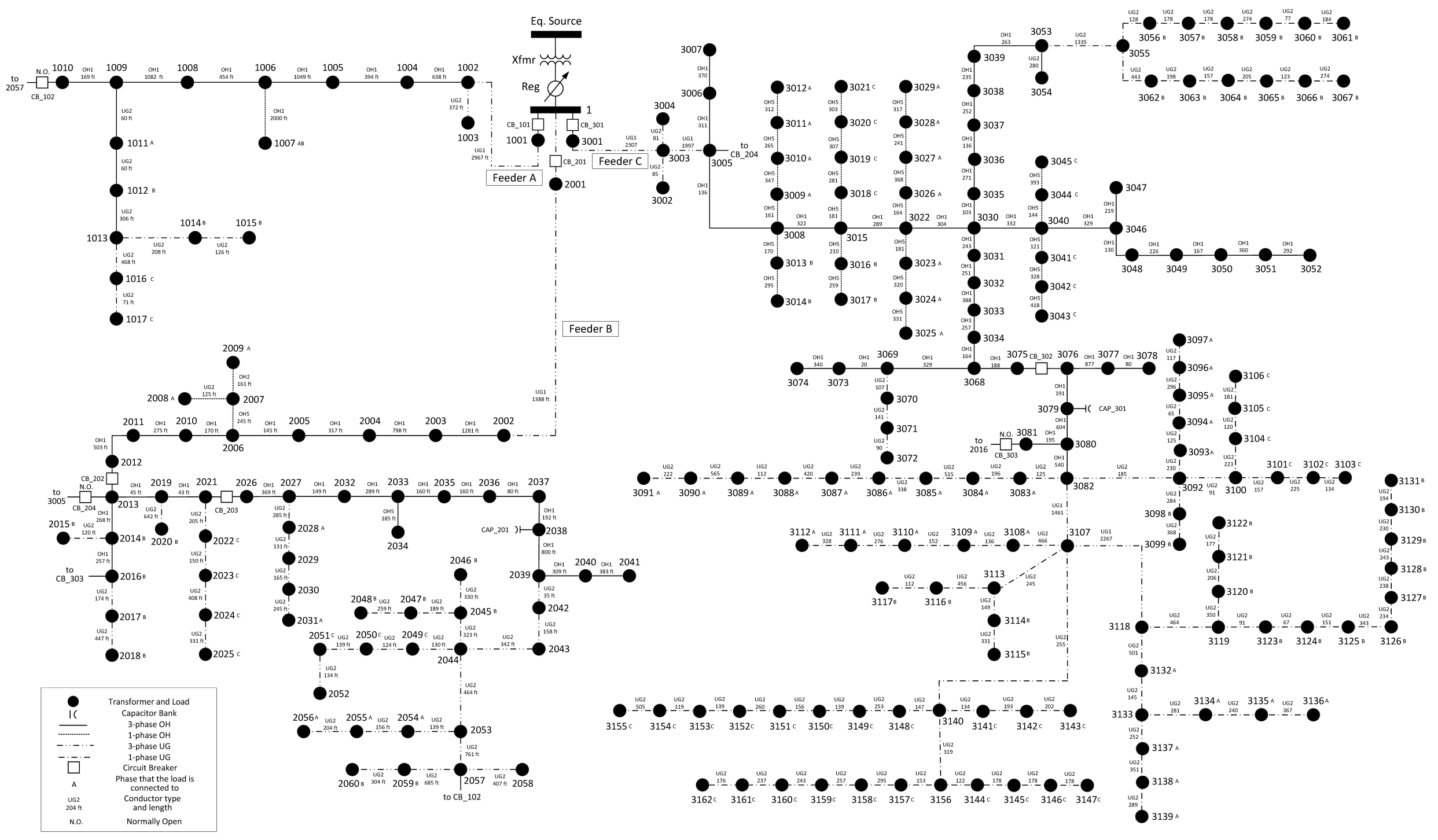}
\caption{Real distribution grid located in the Midwest U.S. \cite{IowaState}, with 240 nodes excluding the substation node.}
\label{fig:grid}
\end{figure}

The RL agent interacts with the distribution grid every 10 ms (the time step), and each episode contains 100 time steps, for a total of one second per episode. $\mu$ is set to $0.1$ to favor voltage regulation over solar production maximization. We use the distribution grid shown in Fig. \ref{fig:grid}, where $N=240$, and we select $n=16$ and $n=194$ for the case studies that follow. 194 is the number of controllable nodes provided originally with the OpenDSS model of this grid. For each of these 194 nodes, we have one year \inserted{(2017)} of real historical load data $(P^l,~Q^l)$, which we take advantage of to generate random samples for our simulation at the start of every episode. \inserted{A short summary of the load data is shown in Fig. \ref{fig:load_stat}.}

Since each episode is 1 second, it is fair to assume that fluctuations in $p^{env}$ and $(P^l,~Q^l)$ are negligible within one episode. For this reason,
% \removed{during the training/learning process, a \textit{reset} command is issued at the beginning of each episode to}
\inserted{at the beginning of each episode, as indicated in Algorithm \ref{alg:algo_episode}, we} randomly generate and fix $p^{env}$ and $(P^l,~Q^l)$ for the remainder of the episode. \inserted{Maximum solar power output, $p^{env}$, for training purposes is randomly selected in each episode as a multiple of the net load at the same bus. For example, at a given bus, if the net load is $x$ kW, then $p^{env}$ at the same bus is chosen uniformly between 0 and $2x$ kW. This helps generate a diverse set of scenarios, ranging from net over-consumption to net over-production at each bus.}

\inserted{The implementation of each episode is detailed in Algorithm \ref{alg:algo_episode}. If the episode is being run for the purpose of training, experiences are collected and stored in the experience buffer (also known as the replay buffer). Furthermore, each agent acts deterministically during online interaction with the environment. However, during training, actions are sampled from a Gaussian, whose mean and variance are determined by the policy's neural network. The \texttt{in\_training} boolean (True or False) is used in the algorithm to distinguish between online execution only, or training. Finally, Algorithm \ref{alg:algo_train} describes how interaction with the environment is used to train the policy and value networks ($\pi_i~\forall i$ and $\bar{V}$).}

\subsection{Case Study on a smaller (16-bus) subsystem}
In this case study, we compare the use of a centralized policy to that of a decentralized policy, presented in Section \ref{policy_optimization}.

Since $n=16$, neural networks of both centralized and decentralized policies have 32 inputs and 32 outputs. The standard choice of 2 hidden layers with 64 neurons per layer is made for the centralized policy, with $tanh(\cdot)$ activation functions, whereas the decentralized policy splits into 16 sub-policies each with 2 inputs, 2 outputs, and two hidden layers each with 4 neurons. This gives both the centralized and decentralized policies a `height' of 64 neurons in the hidden layer ($4\times16 = 64$), but a total of 8352 parameters to tune for the former and 672 for the latter. In fact, in the decentralized case, we assign 16 different Adam optimizers, one to tune each sub-policy, so it's not so much 672 parameters to optimize per PPO iteration, rather 42 per optimizer, compared to 8352 per (single) optimizer in the centralized setting.

\begin{figure}[!ht]
\centering
\includegraphics[width=0.45\textwidth]{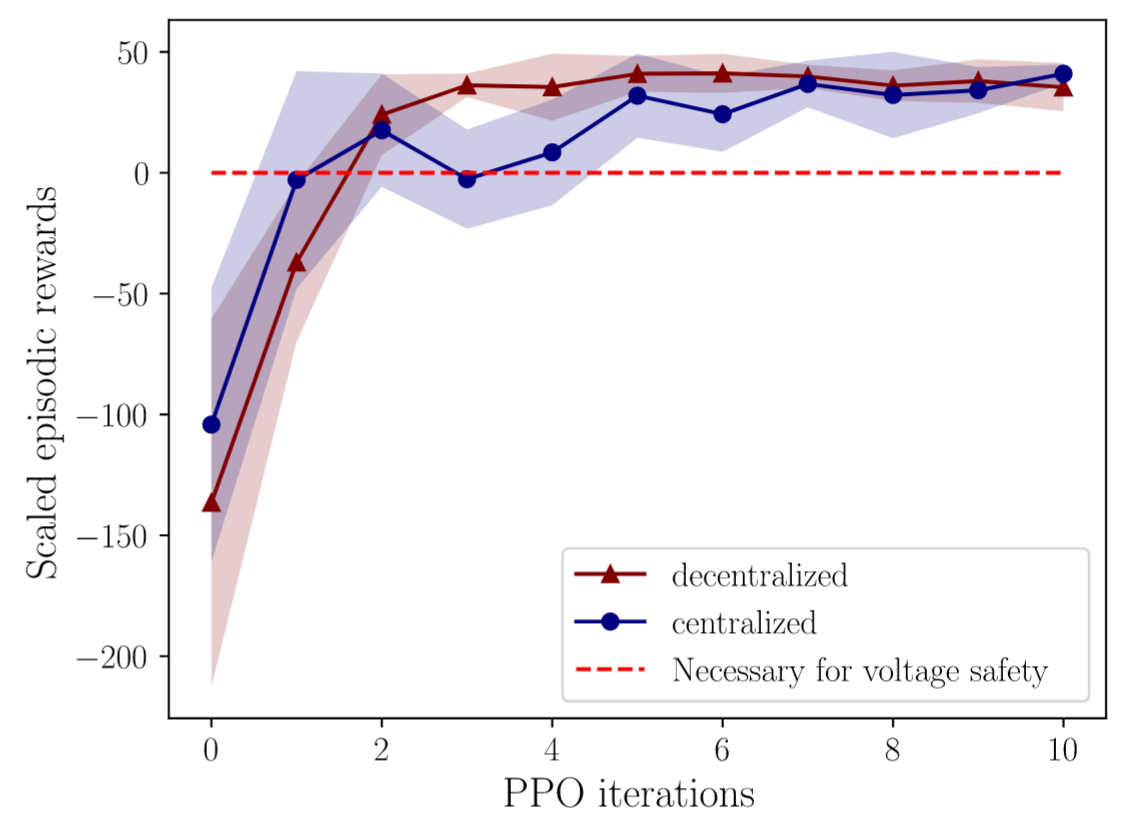}
\caption{RL training curve for centralized vs. decentralized policies. Decentralized agent trains more monotonically and with less variance in episodic rewards. For both, rewards below 0 indicate voltages are not within user-defined safety boundary.}
\label{fig:rewards}
\end{figure}

The training curve for each is shown in Fig. \ref{fig:rewards}, where each `PPO iteration' on the $x$-axis refers to 2048 steps of interacting with the environment (or 20s, considering 10ms time step). It is evident that the centralized agent does not out-perform the decentralized agent, and is clearly less interpretable, and requires a wide communication infrastructure to implement in practice. Note: in both centralized and decentralized cases, \textit{value} function $V$ is centralized (fully connected neural network). That is, the RL agent is centralized during training (computer simulation), but decentralized during execution (real-world).

In classic RL benchmarks, a threshold on average cumulative rewards is chosen to determine when the learning problem is \textit{solved}. This helps the person who monitors and debugs the learning process to get a sense of progress to save time and effort. In our context, the threshold is set to $0$, as shown in Fig. \ref{fig:rewards}, for the following reasons. We know that $R_V\leq0$ and $R_P\geq0$, from Eq. (\ref{eq:R_P},\ref{eq:R_V}). Both reward terms have been designed in such a way that the magnitudes of the rewards are of order 1 or less during normal operating conditions. Moreover, say that the user desires to keep voltages within $1\pm\delta$. It is then a fact that $(R_V+\mu R_P)\geq0$ at every bus only if voltages are kept within the desired region at all buses. It logically follows that if the inequality does not hold, then the voltage at least at one bus must be outside the desired region. Thus, we can state that by simply monitoring the training curve, one may claim that not all voltages are inside $1\pm\delta$ if the curve is still below the threshold, and that certainly more time needs to be given before ending the training progress. This necessary condition on voltage serves as a useful tool for users who seek to implement this approach. \textit{Note:} in that figure, the term 'voltage safety' merely refers to voltages being inside $1\pm\delta$. Fig. \ref{fig:rewards} demonstrates that the decentralized agent permanently crosses this threshold after 2 iterations, while the centralized takes 4 iterations to do so.

By these results, we claim that one can obtain results for a decentralized agent that are similar to, or even better than, those for a centralized agent, simply by manipulating the neural network's architecture.

\subsection{Case Study on a larger (194-bus) subsystem}
In the previous subsection, we compared centralized and decentralized policy architectures. In this subsection, we dig deeper to examine our proposed framework from purely a power systems perspective. We ask the following question: what is the impact of joint real and reactive power control (as opposed to just the latter) on system-wide voltage profile in the midst of deep photovoltaic penetration?

\begin{figure}[!ht]
\centering
\includegraphics[width=0.42\textwidth]{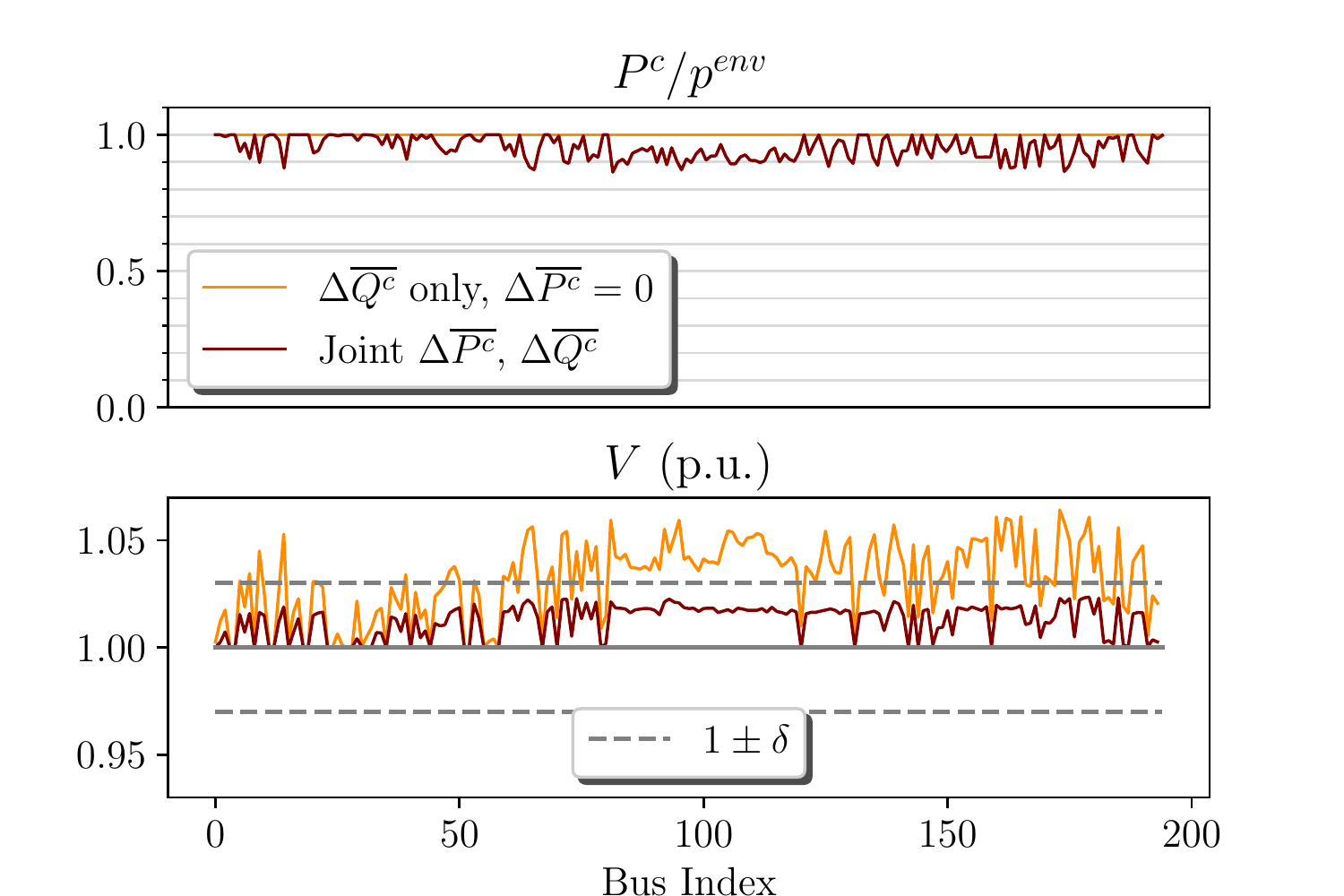}
\caption{\inserted{Comparison between control policies under a scenario with deep photovoltaic penetration. Joint provision of real and reactive power yields improved voltage profile with minor power sacrifice.}}
\label{fig:compare_PV}
\end{figure}

\begin{figure}[!ht]
\centering
\includegraphics[width=0.42\textwidth]{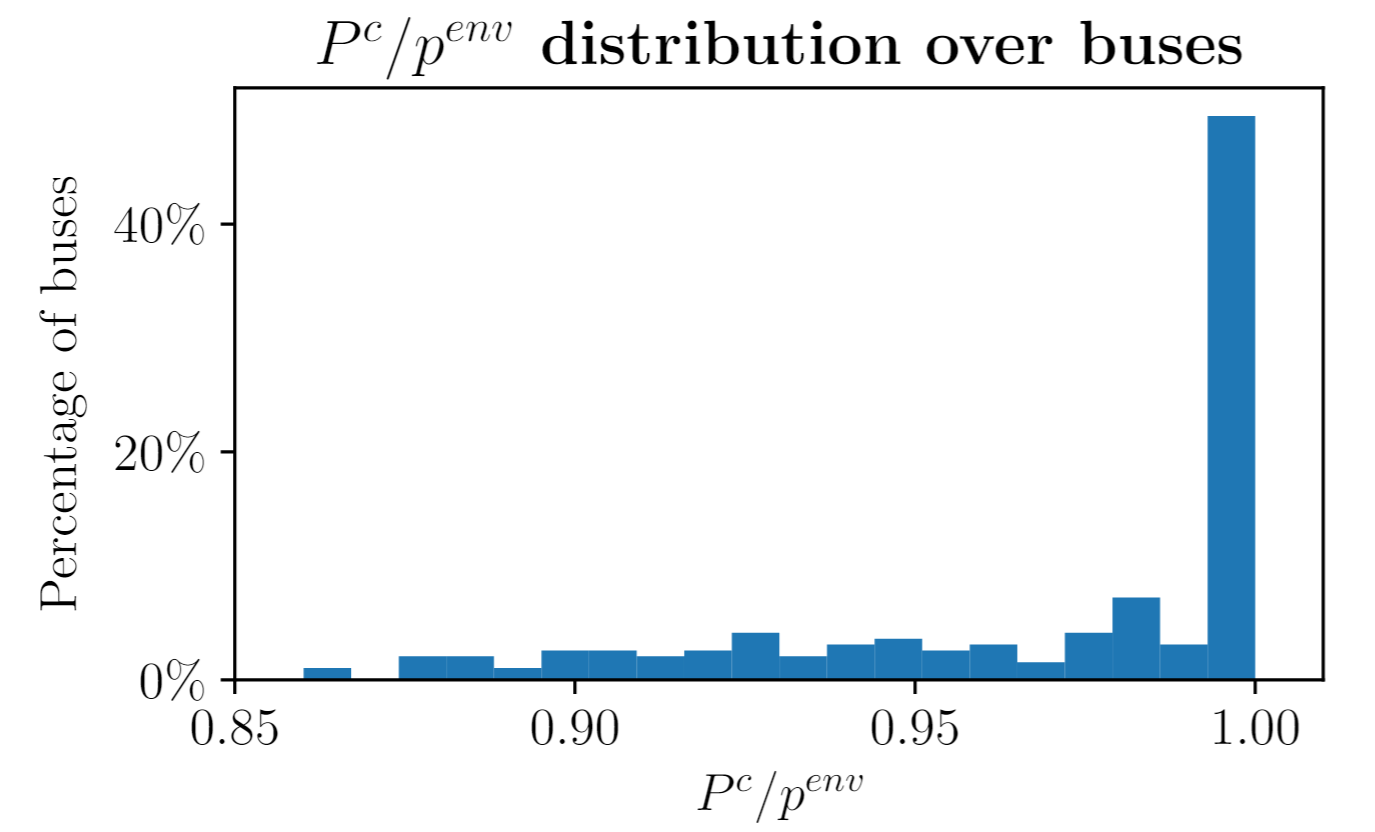}
\caption{Histogram of $P^c/p^{env}$ under RL policy, using results of Fig. \ref{fig:compare_PV}. Most buses inject near-maximum real power.}
\label{fig:Pp_distribution}
\end{figure}

Consider $n=194$ buses, and the same grid as before, with controllable real and reactive power inverter setpoints. As shown in Fig. \ref{fig:compare_PV}, when maximum real power is drawn from the solar panels, leaving less reactive power support, deep photovoltaic penetration causes over-voltage.
% \removed{Terms \textit{Proportional Reactive} and \textit{Integral Reactive} refer to conventional droop control schemes where maximum power is injected and whatever remains within inverter limits is used for reactive power compensation to regulate voltage. The first term refers to reactive power output proportional to voltage by a droop curve (with saturation as usual), while the second is the same but for incremental changes in reactive power.}
With joint provision of real and reactive power, the RL agent manages to keep voltages within the user-defined desired region ($1\pm\delta$). Surprisingly, a small reduction in real power injection was needed to achieve this effect. Fig. \ref{fig:Pp_distribution} shows the steady-state distribution of real power consumption per bus, as a ratio to maximum possible injection ($p^{env}$). It is worth noting how well the voltage was improved system-wide, even though most solar panels produced near maximum output (note the 0.85 on the y-axis of both figures \ref{fig:compare_PV} and \ref{fig:Pp_distribution}). This justifies the value in considering joint provision of real and reactive power support.

\inserted{Indeed, there exist plenty of studies in the body of literature that implement some form of power curtailment (to provide reactive power support) for the purposes of voltage regulation. We propose the novelty here is twofold. Firstly, by using RL to automatically learn to balance between voltage regulation and maximum power utilization, the user does not need to design heuristics to reach similar results or to explicitly rely on any understanding of how the system works. Secondly, as shown in Fig. \ref{fig:compare_PV}, very little power curtailment is required to significantly improve the voltage profile and keep it well within the desired thresholds. This is made possible by the structure of the proposed decentralized PPO policy which allows all the agents to simultaneously train to adapt to one another during exploration, despite the lack of communication between them during real-time operation.}

%%%%%%%%%%%%%%%%%%%%%%%%%%%%%%%%%%
% CONCLUSION
%%%%%%%%%%%%%%%%%%%%%%%%%%%%%%%%%%
\section{Concluding Remarks}
This paper introduces a reinforcement learning-based voltage control strategy with joint provision of real and reactive power 
% \removed{support}
for distribution grids with  deep photovoltaic penetration. The 
% \removed{voltage regulation problem is posed}
\inserted{joint real power and voltage support problem is formulated} as a Markov Decision Process with rewards parametrized to balance between voltage deviation minimization and solar production maximization.

\inserted{Compared with conventional multi-agent RL algorithms, we develop a tailor-designed \textit{decentralized PPO (Proximal Policy Optimization)} algorithm that would both work well with large continuous action spaces and simplify the optimization process by updating multiple policies simultaneously. We demonstrate that by reducing the search space, for the simper 16-bus subsystem, from over 8000 parameters for a centralized setting to under 700 parameters for a decentralized setting, we still achieve similar, and in some cases better, average rewards. This size reduction is on the order of 10 or so, but for the 194-bus subsystem, it is on the order of 100.} Numerical simulations on a 240-node distribution grid based on real parameters show that it is not always the best strategy to absorb all the solar power available. This observation implies that it would benefit the distribution grid the most if some fraction of the real-time power produced from the PV panels can be locally absorbed.

This paper also proposes and verifies a fully decentralized (communication-free) approach for this type of control, which can be implemented on existing physical infrastructure, helping alleviate problems related to communication failure or cyber-attacks. In future work, competition between agents is considered, whereby the inverter at each bus seeks to maximize local, not system-wide, rewards. Further research could also investigate the optimal combination of local energy storage together with PV panels for real-time operation.

% or


\appendices
\section{Experiment Details}
\label{Appendix_Experiment_Details}

\inserted{Software:} Simulations are conducted in Python 3.7, interfacing with OpenDSS \cite{OpenDSS} (unbalanced distribution grid simulator) and using PyTorch \cite{PyTorch} (Python-based library to train neural networks) to model, build and train actor and critic neural networks. 
% \removed{Machine:}
\inserted{Hardware:} Lenovo, 64-bit Windows 10, Intel\textregistered Core\texttrademark i7-6700HQ CPU @ 2.60Ghz, 16.0 GB RAM.

\inserted{Proximal policy optimization (PPO) parameters: Number of steps per training update of 512 to 2048 time steps. Considering each episode/scenario involved 100 time steps, this implies on the order of 10 episodes every actor and critic network update. Batch size of 16 time steps with 10 epochs. This means that for each training update, the aforementioned neural networks are updated by performing nonlinear regression 10 times, each using 16 time steps randomly selected from the experience buffer (refer to Algorithm \ref{alg:algo_train}).}


% \section*{Acknowledgment}

% \section*{Reference Placeholders}
% Temporary dummy section only.

\ifCLASSOPTIONcaptionsoff
  \newpage
\fi


%%%%%%%%%%%%%%%%%%%%%%%%%%%%%%%%%%
% REFERENCES
%%%%%%%%%%%%%%%%%%%%%%%%%%%%%%%%%%
\bibliographystyle{IEEEtran}
\bibliography{IEEEabrv,MAIN.bbl}

%%%%%%%%%%%%%%%%%%%%%%%%%%%%%%%%%%
% BIOGRAPHIES

% \begin{IEEEbiography}{Rayan El Helou}
% \end{IEEEbiography}

% \begin{IEEEbiography}{Dileep Kalathil}
% \end{IEEEbiography}

% % insert where needed to balance the two columns on the last page with
% %\newpage

% \begin{IEEEbiography}{Le Xie}
% \end{IEEEbiography}


%\vfill


\end{document}